\documentclass{ws-mplb}
\usepackage[super]{cite}
\usepackage{hyperref}
\usepackage{url}

\begin{document}

\markboth{Xin Huang and Shuai Dong}{Ferroelectric Control of Magnetism and Transport in Oxide Heterostructures}

%
\catchline{}{}{}{}{}
%

\title{Ferroelectric Control of Magnetism and Transport in Oxide Heterostructures}

\author{Xin Huang and Shuai Dong\footnote{All correspondence should be addressed to S.D.: sdong@seu.edu.cn}}
\address{Department of Physics, Southeast University, Nanjing 211189, China}

\maketitle

\begin{history}
\received{(28 July 2014)}
\end{history}

\begin{abstract}
Magnetism and transport are two key functional ingredients in modern electronic devices. In oxide heterostructures, ferroelectricity can provide a new route to control these two properties via electrical operations, which is scientifically interesting and technologically important. In this Brief Review, we will introduce recent progresses on this fast developing research field. Several subtopics will be covered. First, the ferroelectric polarization tuning of interfacial magnetism will be introduced, which includes the tuning of magnetization, easy axis, magnetic phases, as well as exchange bias. Second, the ferroelectric polarization tuning of transverse and tunneling transport will be reviewed.
\end{abstract}

\keywords{ferroelectricity; oxide heterostructures; ferroelectric field effect; tunnel junctions.}

\graphicspath{{Figs/}}

\section{Introduction}
In modern electronic devices, magnetism and transport are two key functional properties. The information data are mostly stored as magnetic bits, while these information bits are operated in semiconductor junctions with the on/off bistable transport. There are several bottlenecks of these electronic devices. For example, although the giant magnetoresistive effect and tunneling magnetoresistive effect have been successfully applied in magnetic sensors to read magnetic bits, the writing of magnetic bits still needs an electrical coil to generate a magnetic field, which limits the further increasing of storage density and is not power-saving. Further, all semiconductor junctions are electrical volatile, consuming too much energy and heating devices.

Magnetoelectric effect, which denotes the mutual coupling and cross control between electric and magnetic orders, may provide an alternative way to utilize the magnetic or transport properties of materials faster and more efficiently. A straightforward expectation is to record non-volatile magnetic bits using pure electric voltage instead of electric current, which can make devices faster, smaller, more stable, and energy-saving comparing with current ones.

Researches on magnetoelectricity took off more than half a century ago. The magnetoelectric coupling was first predicted to occur in Cr$_2$O$_3$ in $1959$\cite{Dzyaloshinskii:Spj} and was experimentally observed in the early $1960$s.\cite{Astrov:Spj,Folen:Prl} However, the observed magnetoelectric coupling coefficient was very small, only about $10^{-14}$ Oecm$^2$V$^{-1}$. Other single-phase magnetoelectric materials were subsequently discovered to show linear magnetoelectric coupling, such as Ti$_2$O$_3$,\cite{Alshin:Spj} GaFeO$_3$.\cite{Rado:Prl} Almost all these materials show very poor performance on magnetoelectric coupling.

Since $2003$, the breakthrough led by BiFeO$_3$ and TbMnO$_3$, has greatly pushed forward the studies of single-phase multiferroic materials.\cite{Wang:Sci,Kimura:Nat} Even though, till now, few of them, except BiFeO$_3$, has be utilized to fabricate devices, mostly due to their poor performance. An unavoidable physical issue is that usually magnetism and ferroelectricity are mutually excluded.\cite{Hill:Jpcb}

As an alternative way, larger magnetoelectric coupling could be expected in those composite systems involving strong ferroelectric and ferromagnetic materials. Indeed, the greatly enhanced magnetoelectric effect have been discovered experimentally in some two-phase nano-composites, such as BaTiO$_3$-CoFe$_2$O$_4$,\cite{Zheng:Sci} PbTiO$_3$-CoFe$_2$O$_4$,\cite{Li:Apl} BiFeO$_3$-CoFe$_2$O$_4$.\cite{Zheng:Am,Zavaliche:Nl} The magnetoelectric coefficient in these composites can each $10^{-11}$ Oecm$^2$V$^{-1}$, which is about three orders of magnitude higher than those aforementioned single phase magnetoelectric materials. The strong magnetoelectric coupling in these heterostructures originates from the collective interaction between the piezoelectric nature of the ferroelectric phase (BaTiO$_3$, PbTiO$_3$ and BiFeO$_3$) and the magnetostrictive nature of the ferromagnetic phase (CoFe$_2$O$_4$). The similar mechanism has also been explored in the epitaxial La$_{\frac{2}{3}}$Sr$_{\frac{1}{3}}$MnO$_3$/BaTiO$_3$,\cite{Eerenstein:Nm} La$_{\frac{3}{4}}$Sr$_{\frac{1}{4}}$MnO$_3$/PMN-PT\cite{Zheng:Prb1} and Pr$_{0.5}$Ca$_{0.5}$MnO$_3$/PMN-PT\cite{Zheng:Prb2} heterostructures. Such a strain-mediated magnetoelectric coupling even presents in macroscopic multilayers connected using gluewater.\cite{Yu:Apl} In the present review, we will not introduce more on this strain-mediated magnetoelectric coupling. Readers can find more details about magnetoelectric composites in an excellent review by Nan \textit{et al.}.\cite{Nan:Jap}

In additional to the intrinsic magnetoelectric effects in single-phase multiferroics and strain-mediated magnetoelectric coupling in two-phase composites, the electronically driven interfacial magnetoelectric effects have also recently attracted significant research interests for their novel physical mechanism and promising applications. In this Brief Review, we will focus on these interfacial magnetoelectric effects. Some recent progresses on ferroelectric control of magnetism and transport in oxide heterostructures will be introduced in the following.

\section{Ferroelectric polarization tuning of interfacial magnetism}

\subsection{Surface magnetization from screening charge}
From the viewpoint of symmetry, the magnetoelectricity breaks both the space-inversion and time-reversal symmetry at the same time. Based on this point, ferromagnetic interfaces are promising candidates for realizing interfacial magnetoelectric effects, because the time-reversal symmetry is broken by the ferromagnetism and the space-inversion symmetry is violated by the interface.

An early theoretical study by Duan \textit{et al.}\cite{Duan:Prl1} revealed the surface magnetoelectric effect in pure ferromagnetic Fe ($001$), Ni ($001$), and Co ($001$) films. The effect originates from the spin-dependent screening of the electric field which leads to notable changes of the surface magnetization as well as the surface magnetocrystalline anisotropy. The charge screening is a normal effect in any metals. However, when the screened metal is ferromagnetic, the screening charges are spin dependent owing to the exchange splitting, namely, the majority- and minority-spin electrons show different response to the external electric field, which induce an additional surface magnetization. In order to quantify this magnetoelectric effect, a surface magnetoelectric coefficient $\alpha_s$ was introduced as:
\begin{equation}
\alpha_s = \frac{\mu_{\rm B}}{ec^2}\frac{n_\uparrow-n_\downarrow}{n_\uparrow+n_\downarrow},
\label{1}
\end{equation}
where $n_\uparrow$ and $n_\downarrow$ are the surface spin-dependent density of states at the Fermi energy. A linear contribution to the surface magnetization $M_{\rm surf}$ can be determined by $\alpha_s$ as follows:
\begin{equation}
\mu_0M_{\rm surf} = \alpha_sE.
\label{2}
\end{equation}
Following the Eq.~\ref{1}-\ref{2}, the estimated magnetoelectric coefficients for Fe, Co and Ni thin films were found to be very small, $\sim10^{-14}$ Oecm$^2$V$^{-1}$, in the same order as for Cr$_2$O$_3$. In addition, the predicted magnetoelectric coupling is confined to the surface since the additional spin moments mainly dominate only at the surface of films.

The magnetoelectric effect can be substantially magnified at the interface between a ferromagnetic film and a high-$\kappa$ dielectric material. For a given electric field, the screening charge in the metal is proportional to the dielectric constant of adjacent dielectric material. Rondinelli \textit{et al.}\cite{Rondinelli:Nn} predicted a higher magnetoelectric coefficient in the SrRuO$_3$/SrTiO$_3$ heterostructure. The calculated value of $\alpha_s$ is about $10^{-12}$ Oecm$^2$V$^{-1}$, which is higher by two orders of magnitude than the aforementioned Fe films. The enhanced magnetoelectric effect is related to the high-$\kappa$ dielectric SrTiO$_3$.

\begin{figure}
\centering
\includegraphics[width=\textwidth]{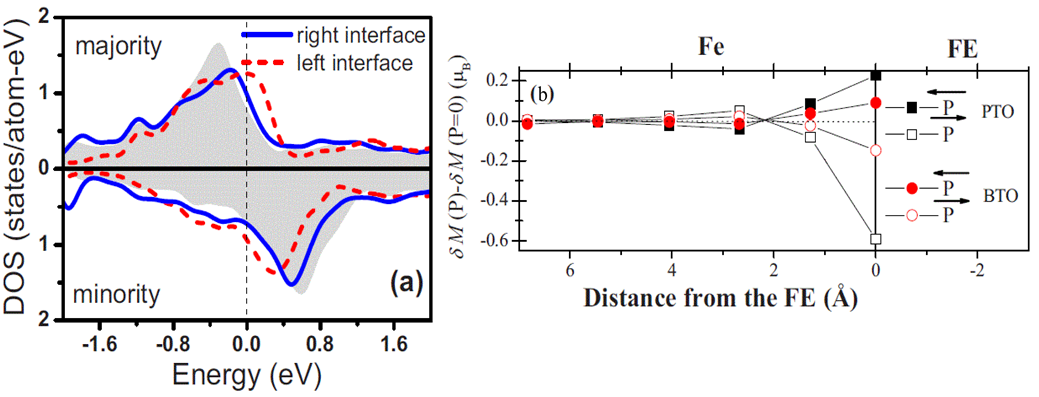}
\caption{(a) Spin-polarized local density of states projected onto Ru's $3d$ orbitals at the right (solid lines) and left (dashed lines) interfaces in the SrRuO$_3$/BaTiO$_3$ ($001$) heterostructure. The shaded plots are the Ru's $3d$ orbital density of states in the bulk. Reprinted figure with permission granted from M. K. Niranjan \textit{et al.}, \textit{Appl. Phys. Lett.} \textbf{95} (2009) 052501. Copyright \copyright (2009) by the American Institute of Physical, \href{http://dx.doi.org/10.1063/1.3193679}{http://dx.doi.org/10.1063/1.3193679}. (b) The magnitude of polarization-induced-magnetization for two Fe atoms per lateral unit cell \textit{vs} distance from the Fe/ferroelectric interface in the PbTiO$_3$ (PTO) and BaTiO$_3$ (BTO) based superlattices. Reprinted figure with permission granted from J. Lee \textit{et al.}, \textit{Phys. Rev. B} \textbf{81} (2010) 144425. Copyright \copyright (2010) by the American Physical Society, \href{http://dx.doi.org/10.1103/PhysRevB.81.144425}{http://dx.doi.org/10.1103/PhysRevB.81.144425}.}
\label{f1}
\end{figure}

This electronically driven magnetoelectric effect can be further enhanced by in ferromagnet/ferroelectric heterostructures. The ferroelectric polarization is equivalent to surface charge at the interface. The surface charge density can be estimated from the value of polarization, e.g. $10$ $\mu$C/cm$^2$ corresponds to $0.1$ electron per unit cell area if the qseudo-cubic lattice constant is $\sim4$ angstrom. In this sense, the majority-spin and minority-spin densities of states at the interface can be significantly changed to different levels by switching the ferroelectric polarization's orientation. Such a magnetoelectric effect was predicted in the SrRuO$_3$/BaTiO$_3$/SrRuO$_3$ heterostructure.\cite{Niranjan:Apl1} As a result of the ferroelectricity in BaTiO$_3$, the magnetization of SrRuO$_3$ at two interfaces differs significantly. Fig.~\ref{f1}(a) shows the spin-polarized density of states projected onto Ru's $3d$ orbitals at the right and left interfaces. It is clear that the exchange splitting is changed between these two interfaces, giving rise to a modulation in the electron population of the two spin channels. The obtained change of magnetic moment caused by the ferroelectric polarization reversal is $0.31$ $\mu_B$ per Ru per interface. Hence, the surface magnetoelectric coefficient can be estimated using Eq.~\ref{2} by assuming a typical coercive field of BaTiO$_3$, which leads to a large $\alpha_s\sim10^{-10}$ Oecm$^2$V$^{-1}$. The calculation also shows that the magnetoelectric coupling in this system displays a highly nonlinear dependence on the magnitude of the ferroelectric polarization, which is different from the linear effect in the aforementioned non-ferroelectric cases.\cite{Duan:Prl1,Rondinelli:Nn}

Meanwhile, Cai \textit{et al.}\cite{Cai:Prb} proposed a strategy to realize the room-temperature magnetoelectric effect in a tri-component ferromagnet/ferroelectric/normal-metal superlattice. Due to the broken inversion symmetry between the ferromagnet/ferroelectric and normal-metal/ferroelectric interfaces, the additional magnetization caused by spin-dependent screening will accumulate at the ferromagnet/ferroelectric interface and not be canceled by the depletion at the normal-metal/ferroelectric interface. Therefore, large global magnetization can be induced in this superlattice. Based on this model, Lee \textit{et al.} predicted a robust magnetoelectric coupling in the Fe/BaTiO$_3$/Pt and Fe/PbTiO$_3$/Pt superlattices.\cite{Lee:Prb} Fig.~\ref{f1}(b) shows the polarization-induced-magnetization of Fe near the BaTiO$_3$ and PbTiO$_3$ interfaces for the positive/negative ferroelectric polarizations. The change of magnetic moments by switching the ferroelectric direction is significant and the induced moments decay rapidly away from the interface. The induced interfacial magnetization is much larger when using PbTiO$_3$ than BaTiO$_3$ since the former has a larger spontaneous polarization. The estimated $\alpha_s$ for Fe/BaTiO$_3$/Pt reaches $10^{-11}$ Oecm$^2$V$^{-1}$, which increases up to $10^{-9}$ Oecm$^2$V$^{-1}$ for Fe/PbTiO$_3$/Pt. The magnitude of this charge-mediated magnetoelectric coupling exceeds the strain-mediated magnetoelectric susceptibility observed in CoFe$_2$O$_4$/BiFeO$_3$.\cite{Zavaliche:Nl}

\subsection{Interfacial bonding tuning magnetization}
\begin{figure}
\centering
\includegraphics[width=\textwidth]{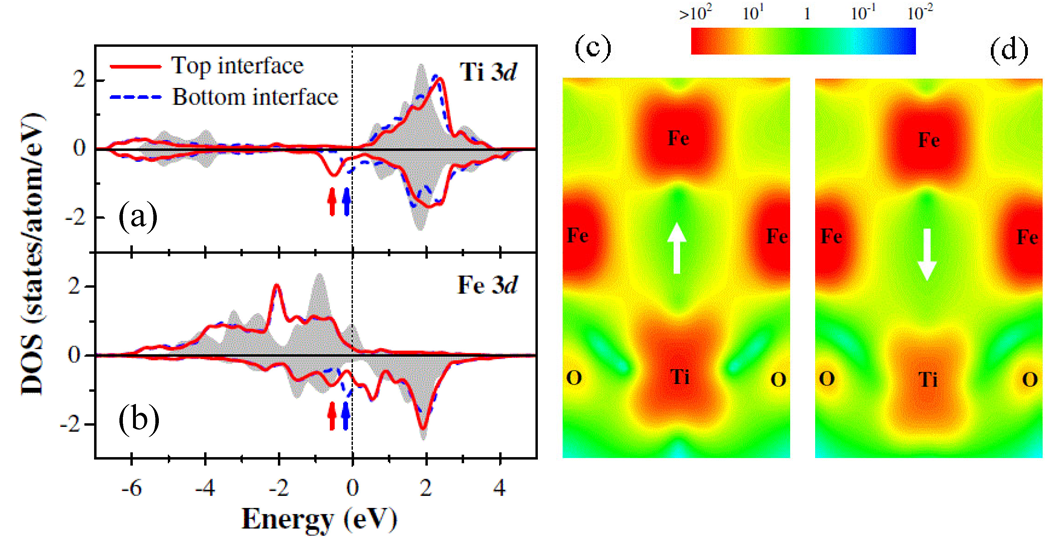}
\caption{Orbital-resolved density of states for interfacial atoms in the Fe/BaTiO$_3$ multilayers: (a) Ti's $3d$, (b) Fe's $3d$. Majority- and minority-spin density of states are shown in the upper and lower panels, respectively. The shaded plots are the density of states of corresponding atoms in the central monolayer which can be regarded as in bulk. (c-d) Minority-spin charge density at the Fe/BaTiO$_3$ interface for two opposite polarizations of BaTiO$_3$. (c) Polarization up. (d) Polarization down. Reprinted figure with permission granted from C.-G. Duan \textit{et al.}, \textit{Phys. Rev. Lett.} \textbf{97} (2006) 047201. Copyright \copyright (2006) by the American Physical Society, \href{http://dx.doi.org/10.1103/PhysRevLett.97.047201}{http://dx.doi.org/10.1103/PhysRevLett.97.047201}.}
\label{f2}
\end{figure}

Besides above simple mechanism of spin-dependent screening, the interfacial bonding may also play a crucial role for the magnetoelectric effect at the ferromagnet/ferroelectric interfaces. For example, Duan \textit{et al.}\cite{Duan:Prl2} studied Fe/BaTiO$_3$ multilayers as a model system. Their calculations show a large magnetoelectric effect at the interface and the hybridization between Ti's $3d$ and Fe's $3d$ states at the surface should be responsible for this effect. Fig.~\ref{f2}(a)-(b) show the orbital-resolved local density of states for Ti's $3d$ and Fe's $3d$ bands. It is clearly that Ti's $3d$ bands hybridize strongly with Fe's minority-spin $3d$ bands around the Fermi level. This strong hybridization leads to the formation of bonding states, which are pushed down in energy and peaked just below the Fermi level as indicated by arrows in Fig.~\ref{f2}(a)-(b). The bonding states cause a larger occupation of Ti's minority-spin bands than that of the majority-spin ones, which results in a net magnetic moment on Ti. The induced magnetic moment on Ti site is sensitive to the interfacial bonding strength and can be controlled by switching the ferroelectric polarization of BaTiO$_3$, as shown in Fig.~\ref{f2}(c)-(d). When the polarization of BaTiO$_3$ points toward the interface, the Fe-Ti bond length becomes shorter, which strengthens the bonding coupling and thus pushes the minority-spin bonding states down to even lower in energy. As a result, the Ti's minority-spin bands become more populated, giving rise to a large magnetic moment on Ti. While the polarization points away from the interface, the Fe-Ti bond length increases, and thus weakens the bonding coupling. Then the minority-spin bonding states lie at higher energy, so it is less populated and the Ti's magnetic moment decreases. The difference of Ti's magnetic moments for these two opposite polarization directions reaches $0.22$ $\mu_B$, so the value of $\alpha_s$ is estimated to be $10^{-11}$ Oecm$^2$V$^{-1}$, which is on the same order of magnitude as measured in those epitaxial nanostructures.\cite{Zavaliche:Nl} First-principles calculations have revealed that such an ``interfacial bonding-driven" magnetoelectric effect also occurs in many other heterostructures, including Fe$_3$O$_4$/BaTiO$_3$,\cite{Niranjan:Prb} Co$_2$MnSi/BaTiO$_3$,\cite{Yamauchi:Apl} Fe/PbTiO$_3$\cite{Fechner:Prb,Fechner:Pssb} and Co/PbZr$_x$Ti$_{1-x}$O$_3$.\cite{Borisov:Prb} Very recently, this predicted large magnetoelectric effect was observed at the Fe/BaTiO$_3$ interface experimentally.\cite{Radaelli:Nc}

\subsection{Tuning of magnetocrystalline anisotropy}
\begin{figure}
\centering
\includegraphics[width=\textwidth]{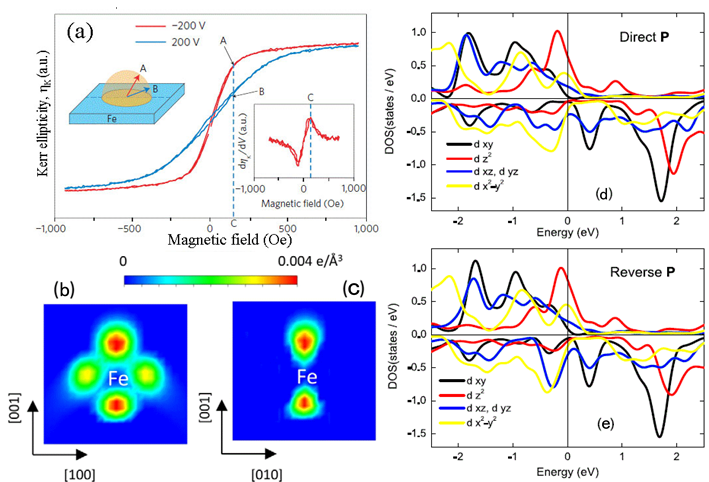}
\caption{(a) Magneto-optical Kerr ellipticity $\eta_k$ for different applied voltages as a function of magnetic field. The right inset shows the voltage modulation response of the Kerr ellipticity, and the left inset illustrates the magnetization direction at points A and B in the hysteresis curves. Reprinted figure with permission granted from T. Maruyama \textit{et al.}, \textit{Nat. Nano.} \textbf{4} (2009) 158. Copyright \copyright (2009) by the Nature Publishing Group, \href{http://dx.doi.org/10.1038/nnano.2008.406}{http://dx.doi.org/10.1038/nnano.2008.406}. (b)-(c) Calculated charge density difference near the Fe atom at the Fe/BaTiO3 interface between the direct and reverse polarization states in the BaTiO$_3$/Fe$_4$/Cu$_3$ ($001$) multilayer system in (b) the ($100$) plane and (c) the ($010$) plane. (d)-(e) Spin-resolved density of states projected to different $d$-orbitals of the interfacial Fe atoms for (d) the direct and (e) reverse polarization states. Reprinted figure with permission granted from P. V. Lukashev \textit{et al.}, \textit{J. Phys.: Condens. Matter} \textbf{24} (2012) 226003. Copyright \copyright (2012) by the Institute of Physics, \href{http://dx.doi.org/10.1088/0953-8984/24/22/226003}{http://dx.doi.org/10.1088/0953-8984/24/22/226003}.} \label{f3}
\end{figure}

The interfacial magnetoelectric effects involve not only electrically-controlled surface (interfacial) magnetization, but also electrically-controlled magnetocrystalline anisotropy. Since the magnetocrystalline anisotropy determines the preferential orientation of magnetization, it may yield entirely new devices for magnetic data storage if the magnetocrystalline anisotropy can be tuned by electric fields. Recently, manipulation of magnetocrystalline anisotropy by applying an electric field has been studied and achieved in many systems with ferromagnetic metallic interfaces (surfaces)\cite{Tsujikawa:Prl,Nakamura:Prl,Duan:Prl1,Niranjan:Apl2,He:Apl,Weisheit:Sci,Maruyama:Nn,Nozaki:Apl} as well as at the interfaces (surfaces) of dilute magnetic semiconductors.\cite{Ohta:Apl,Chiba:Nat} In particular, Maruyama \textit{et al.}\cite{Maruyama:Nn} obtained a significant change in perpendicular magnetocrystalline anisotropy at the Fe/MgO ($001$) interface by an electrical voltage, which reached up to $40\%$ in energy difference for a relatively small applied fields in the order of $100$ mV/nm, as shown in Fig.~\ref{f3}(a). A significant change of the hysteresis loop indicates a remarkable change of the perpendicular magnetic anisotropy following the application of bias voltage. This effect is attributed to the modulation of electron occupation of Fe's $3d$ orbitals adjacent to the Fe/MgO interface. The underlying physical origin of magnetocrystalline anisotropy, namely the spin-orbit coupling, depends on the electronic structure of Fe. Therefore, a strong modification of the surface magnetocrystalline anisotropy can be obtained by tuning the electron density in Fe, as confirmed by first-principles calculations.\cite{Duan:Prl1,Niranjan:Apl2,Nakamura:Prl,Nakamura:Prb} Similar to aforementioned cases, since the electric field is applied across MgO with a large dielectric constant, the magnetoelectric effect can be substantially enhanced.\cite{Rondinelli:Nn} Very recently, electrically induced bistable magnetization switching was realized in magnetic tunnel junctions at room temperature.\cite{Wang:Nm,Shiota:Nm}

Alternatively, the interface magnetocrystalline anisotropy can be more efficiently controlled by a ferroelectric polarization of an adjacent ferroelectric film. This approach may provide a bistable magnetization state directly coupled to the bistable polarization state of the ferroelectric layer. The modulation of magnetocrystalline anisotropy by the ferroelectric polarization was predicted by \emph{ab} \emph{initio} studies.\cite{Duan:Apl,Sahoo:Prb,Choi:Jap,Lee:Jap,Fechner:Prl,Lukashev:Jpcm} For example, Lukashev \textit{et al.}\cite{Lukashev:Jpcm} studied the effect of ferroelectric polarization of BaTiO$_3$ to the magnetic anisotropy of Fe. They considered a more realistic geometry, which consists of BaTiO$_3$/Fe$_4$ multilayer instead of the monolayer Fe used in previous studies.\cite{Duan:Apl,Choi:Jap} They found a large change (about $30$\%) in perpendicular interface magnetic anisotropy energy by reversing the ferroelectric polarization of BaTiO$_3$. The underlying physical mechanism can be understood by analyzing the charge density and spin-orbital-resolved density of states. First, as shown in Fig.~\ref{f3}(b)-(c), a notable anisotropy is observed for the electron accumulation along the [$100$] and [$010$] directions in the ($001$) plane. The induced charge density in the [$100$] direction is larger than that in the [$010$] direction. Second, as shown in Figs.~\ref{f3}(d)-(e), the largest contribution to the induced surface charge is from the $3d$$_{z^2}$ orbitals while the in-plane orbital contribution is relatively small. Comparing Figs.~\ref{f3}(d) and (e), a nearly rigid shift of the majority-spin $3d_{z^2}$'s density of states occurs, implying that the amount of the $3d_{z^2}$ electrons is reduced when the polarization is reversed. Further, this redistribution of electron charge over Fe's $3d$ orbitals at the interface driven by the polarization reversal is responsible to the change of magnetocrystalline anisotropy. As stated before, the magnetocrystalline anisotropy is a result of spin-orbit coupling between the occupied valence bands and unoccupied conducting bands. Thus the occupation of $3d_{z^2}$ orbitals will affect the spin-orbit coupling and magnetocrystalline anisotropy.\cite{Nakamura:Prl,Duan:Prl2}

In addition, other ferroelectric materials, like organic ferroelectrics, have also been adopted to tune the magnetocrystalline anisotropy, such as poly (vinylidene fluoride) PVDF/Co,\cite{Lukashev:An} PVDF/Fe,\cite{Wang:Jap} $70$\% vinylidene fluoride with $30$\% trifluoroethylene copolymer P(VDF-TrEE)/Co/Pd.\cite{Mardana:Nl} In these heterostructures, significant changes in magnetocrystalline anisotropy energy driven by the ferroelectric polarization reversal have been found, some of which even reach up to $50$\%. These results provide a new insight into the mechanism of the magnetoelectric coupling at organic ferroelectric/ferromagnet interfaces.

These cutting-edge works give a significant step towards energy-efficient switching of magnetic elements and demonstrate the capabilities of electrical-control for magnetic data storage applications.

\subsection{Ferroelectric control of magnetic phases}
In addition to above continuous tunings of magnetic properties, a more dramatic magnetoelectric effect is predicted to occur at the surface of a magnetic correlated electron systems, such as manganite La$_{1-x}A_x$MnO$_3$ ($A$ = Ca, Sr, or Ba). It is well known that the correlated electronic materials show rich phase diagrams as a function of doped carrier concentration, which consist of plethoric competing phases with different resistive, structures, as well as magnetic orders.\cite{Tokura:Rpp,Tokura:Bok} In particular, when the doping concentration is close to any phase boundaries, it would be possible to switch between two phases (e.g. ferromagnetic and antiferromagnetic) using an external electric field and therefore induce a gigantic magnetoelectric effect. Although the driven force is also based on the screening charge, the underlying mechanism goes beyond the simple addition of spin-polarized carriers as in the aforementioned simple ferromagnetic metals, which relies on charge-induced modification of the magnetic ground state.

\begin{figure}
\centering
\includegraphics[width=\textwidth]{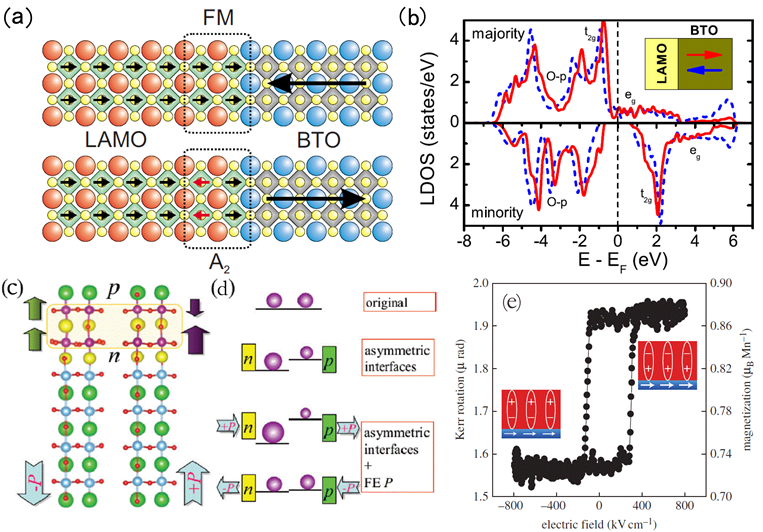}
\caption{(a) Electrically induced magnetic reconstruction at the La$_{1-x}A_x$MnO$_3$/BaTiO$_3$ interface. The interfacial magnetic moments change from the ferromagnetism to A-type antiferromagnetism as the ferroelectric polarization of BaTiO$_3$ is reversed. (b) Spin-resolved local density of states of the interfacial La$_{0.5}A_{0.5}$MnO$_3$ unit cell, with the polarization pointing away from (solid) and toward (dashed) the interface. Reprinted figure with permission granted from J. D. Burton \textit{et al.}, \textit{Phys. Rev. B} \textbf{80} (2009) 174406. Copyright \copyright (2009) by the American Physical Society, \href{http://dx.doi.org/10.1103/PhysRevB.80.174406}{http://dx.doi.org/10.1103/PhysRevB.80.174406}. (c) Sketch of the bilayer design. The $n$-/$p$- type interfaces are indicated. Left/right are the $-P$/$+P$ cases, with switched magnetic orders (ferromagnetic/ferrimagnetic). (d) The $e_{\rm g}$ density (spheres) and potential (bars) modulated by asymmetric interfaces (bricks) and ferroelectric polarization $P$ (arrows). Reprinted figure with permission granted from S. Dong \textit{et al.}, \textit{Phys. Rev. B} \textbf{88} (2013) 140404(R). Copyright \copyright (2013) by the American Physical Society, \href{http://dx.doi.org/10.1103/PhysRevB.88.140404}{http://dx.doi.org/10.1103/PhysRevB.88.140404}. (e) A magnetoelectric hysteresis loop at 100 K shows the magnetic response of the PbZr$_{0.2}$Ti$_{0.8}$O$_3$/La$_{0.8}$Sr$_{0.2}$MnO$_3$ system as a function of the applied electric field. Insets: the magnetic and electric states of the La$_{0.8}$Sr$_{0.2}$MnO$_3$ and PbZr$_{0.2}$Ti$_{0.8}$O$_3$ layers, respectively. Reprinted figure with permission granted from J. A. Molegraaf \textit{et al.}, \textit{Adv. Mater.} \textbf{21} (2009) 3470. Copyright \copyright (2009) by the John Wiley \& Sons, Inc., \href{http://dx.doi.org/10.1002/adma.200900278}{http://dx.doi.org/10.1002/adma.200900278}.}
\label{f4}
\end{figure}

Indeed, a dramatic magnetoelectric coupling has been predicted in La$_{1-x}$A$_x$MnO$_3$/BaTiO$_3$ ($001$) system by the first-principles method.\cite{Burton:Prb} The adopted doping concentration $x$ is $0.5$ in their calculation, chosen to be near the ferromagnetic-antiferromagnetic phase transition.\cite{Tokura:Rpp} The direction of BaTiO$_3$ polarization is used to electrostatically modulate the hole-carrier density in La$_{0.5}A_{0.5}$MnO$_3$ and therefore to modulate the magnetism of La$_{0.5}A_{0.5}$MnO$_3$ between ferromagnetic and antiferromagnetic at the interface, as shown in Fig.~\ref{f4}(a). The origin of this interfacial magnetic reconstruction can be revealed by examining the effects of polarization reversal on the electronic structure at the interface, as shown in Fig.~\ref{f4}(b). When the polarization points away from the interface, there is an apparent upward shift of the local density of states, implying the electron population at the interface is reduced which corresponds to the hole charge accumulation state. The opposite situation occurs when the polarization points to the interface, leading to a downward shift of the local density of states and corresponding to the hole depletion state. Hence, the interface layer acts as a magnetic switch to favor either the antiferromagnetic state (hole accumulation) or the ferromagnetic state (hole depletion) depending on the polarization's orientation, which gives rise to a large change in the magnetic moment and thus a large magnetoelectric effect.

Experimentally, this particular magnetoelectric coupling has been confirmed in the La$_{0.8}$Sr$_{0.2}$MnO$_3$/PbZr$_{0.2}$Ti$_{0.8}$O$_3$ heterostructure.\cite{Molegraaf:Am,Vaz:Prl,Vaz:Apl} Variation of the magnetic hysteresis loop has been observed from the magneto-optical Kerr measurement, responding to the PbZr$_{0.2}$Ti$_{0.8}$O$_3$'s polarization switching, as shown in Fig.~\ref{f4}(e). A larger coercivity and a smaller saturation magnetization are observed for the accumulation state as compared to the depletion one. It is seen that the magnetic reconstruction only occurs within a few atomic layers thickness near the interface, while the rest La$_{0.8}$Sr$_{0.2}$MnO$_3$ film sustains a robust ferromagnetic order during the polarization switch,\cite{Vaz:Jap,Yi:Prl,HLu:Apl,Jiang:Apl} in consistent with the theoretical prediction.\cite{Burton:Prb}

Besides the first-principles approach, a microscopic model based on the two-orbital double-exchange has also been introduced to describe this ferroelectric screening effect in manganites.\cite{Dong:Prb11} The model simulation confirmed that the charge accumulation/depletion near the interface can drive the interfacial phase transition, which gives rise to a robust magnetoelectric response and bipolar resistive switching, in qualitative agreement with those density functional theory calculation.\cite{Burton:Prb} However, the interfacial magnetic phase is different between the density functional theory calculation and model simulation: an uncompensated magnetic interfacial layer is predicted in the former while a compensated one is suggested in the latter. A recent experiment seems to support the compensated interfacial magnetism.\cite{Ma:Apl}

Based on this model, Dong \textit{et al.}\cite{Dong:Prb13} proposed a design to pursuit the full control of magnetism by reversing ferroelectric polarization in a manganite bilayer-ferroelectric superlattice. The design is sketched in Fig.~\ref{f4}(c). The asymmetric polar interfaces: the $n$-type and $p$-type ones, are adopted. The $n$-type interface will attract more electrons to its nearest-neighbour Mn layer, while the $p$-type interface will repel electrons away from the interface. Therefore, even without the external ferroelectricity, the asymmetric interfaces already modulate the electronic density distribution and electrostatic potential within the manganite bilayers. When the ferroelectric polarization points to the $n$-type interface (the $+P$ case), the electrostatic potential difference between the two MnO$_2$ layers is further split, which enhances the charge disproportion. In contrast, the electrostatic potential from the polar interfaces will be partially or fully compensated when the ferroelectric polarization points to the $p$-type interface (the $-P$ case), which suppresses the electronic disproportion, as shown in Fig.~\ref{f4}(d). Due to the largest interface/volume ratio (up to $100$\%) for the bilayer structure, every manganite layer can be fully controlled by the ferroelectric polarization. A remarkable variation in total magnetization (up to $\sim90$\%) tuned by ferroelectric is achieved.

\subsection{Electrical-controllable exchange bias}
The exchange bias effect is known to be associated with the exchange anisotropy created at the interface between ferromagnetic and antiferromagnetic materials,\cite{Nogues:Mmm} where the exchange coupling gives rise to a shift of the magnetic hysteresis loop away from the center of symmetry at zero magnetic field. The exchange bias effect can be understood theoretically as induced by spin pinning effect at the ferromagnetic/antiferromagnetic interface. This effect has been widely used in a variety of magnetic storage and sensor devices. Control of exchange bias using electrical methods would provide a new path to control of magnetization.

The first demonstration of exchange bias controlled by electric field was reported in [Co/Pt]/Cr$_2$O$_3$ ($111$) heterostructures.\cite{Borisov:Prl} The application of an electric field to Cr$_2$O$_3$ can result in a net magnetization whose direction depends on the sign of the electric field. Hence, the direction of exchange bias can be controlled by appropriate electric fields. Subsequently, experimental studies of exchange bias were done on a number of multiferroic materials, such as YMnO$_3$\cite{Dho:Apl,Marti:Apl} and BiFeO$_3$.\cite{Dho:Am} Indeed, Laukhin \textit{et al.}\cite{Laukhin:Prl} observed a large increase of the exchange bias in the NiFe/YMnO$_3$ heterostructure. However, due to the low magnetic ordering temperature of YMnO$_3$, the observed large effect only exists at very low temperatures. Fortunately, BiFeO$_3$ brings hope since its superior properties can be extended to high temperatures. Chu \textit{et al.}\cite{Chu:Nm}and B$\acute{e}$a \textit{et al.}\cite{Bea:Prl} studied the FeCo/BiFeO$_3$ and CoFeB/BiFeO$_3$ heterostructures and found an apparent change in the exchange bias by tuning the electric filed at room temperature, which were suggested to be linked closely with the particular ferroelectric domain wall configuration in BiFeO$_3$.

\begin{figure}
\centering
\includegraphics[width=\textwidth]{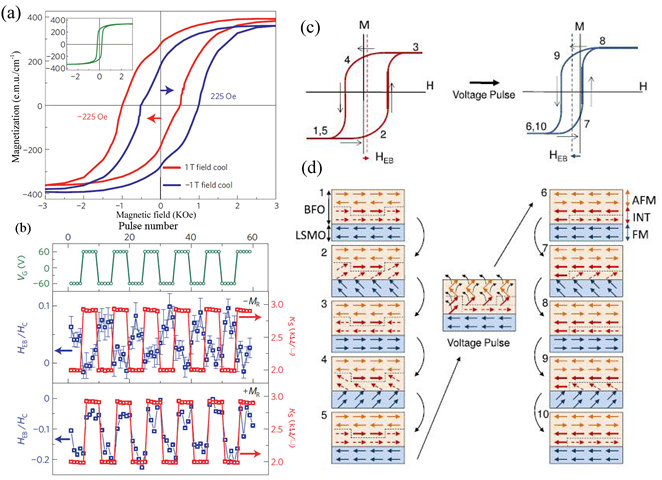}
\caption{(a) Magnetic hysteresis loops of BiFeO$_3$/La$_{0.7}$Sr$_{0.3}$MnO$_3$ heterostructures measured at $7$ K after +/- $1$ T field cooling from $350$ K, respectively. The inset shows a magnetic hysteresis loop for a BiFeO$_3$/SrTiO$_3$/La$_{0.7}$Sr$_{0.3}$MnO$_3$ structure, without an exchange bias after field cooling. (b) Electric-field control of exchange bias. The top panel: the gate-voltage-pulse sequence used for the measurements. The middle and bottom panels: measurement of normalized exchange bias and peak resistance for the gate-pulse sequence shown in the top panel. Reprinted figure with permission granted from S. M. Wu \textit{et al.}, \textit{Nat. Mater.} \textbf{9} (2010) 756. Copyright \copyright (2010) by the Nature Publishing Group, \href{http://dx.doi.org/10.1038/NMAT2803}{http://dx.doi.org/10.1038/NMAT2803}. (c) Magnetic hysteresis loops of La$_{0.7}$Sr$_{0.3}$MnO$_3$ before and after the ferroelectric polarization reversal of BiFeO$_3$. The progression of measurement is denoted by numbers and arrows as magnetic field is swept. (d) Depiction of interfacial spins for each number in (c). Reprinted figure with permission granted from S. M. Wu \textit{et al.}, \textit{Phys. Rev. Lett.} \textbf{110} (2013) 067202. Copyright \copyright (2013) by the American Physical Society, \href{http://dx.doi.org/10.1103/PhysRevLett.110.067202}{http://dx.doi.org/10.1103/PhysRevLett.110.067202}.}  \label{f5}
\end{figure}

This electrical-controllable exchange bias was also found in the BiFeO$_3$/La$_{0.7}$Sr$_{0.3}$MnO$_3$ heterostructures.\cite{Wu:Nm10,Yu:Prl} Fig.~\ref{f5}(a) shows typical magnetic hysteresis loops of BiFeO$_3$/La$_{0.7}$Sr$_{0.3}$MnO$_3$ heterostructures, in which there is a corresponding shift after the field cooling. This exchange bias shift is due to the antiferromagnetic-ferromagnetic coupling since the shift is absent when a thin SrTiO$_3$ layer is inserted between the BiFeO$_3$ and La$_{0.7}$Sr$_{0.3}$MnO$_3$, as shown in insert of Fig.~\ref{f5}(a). More exciting, the observed two distinct exchange-bias states can be operated reversibly by switching the ferroelectric polarization of BiFeO$_3$, as shown in Fig.~\ref{f5}(b). It is clearly that the magnitude of exchange bias is modulated by the electric field between a high value and a low one, which is accompanied with the modulation of channel resistance as well as magnetic coercivity.\cite{Wu:Prl} Fig.~\ref{f5}(c) sketches the progression of the magnetic moments in terms of the hysteresis loops. As the magnetic field is swept [Fig.~\ref{f5}(d)($1$-$5$)], a large number of interfacial spins rotate with the magnetic field while a small fraction moments remain pinned, which results in the high-coercivity and low exchange bias state. After applying a negative voltage pulse to the gate and changing the ferroelectric polarization, the system is now in a different state in which a large fraction of interfacial spins are pinned in the antiferromagnetic layer, which results in the high magnitude of negative exchange bias and low coercivity [Fig.~\ref{f5}(d)($6$)].
\begin{figure}
\centering
\includegraphics[width=\textwidth]{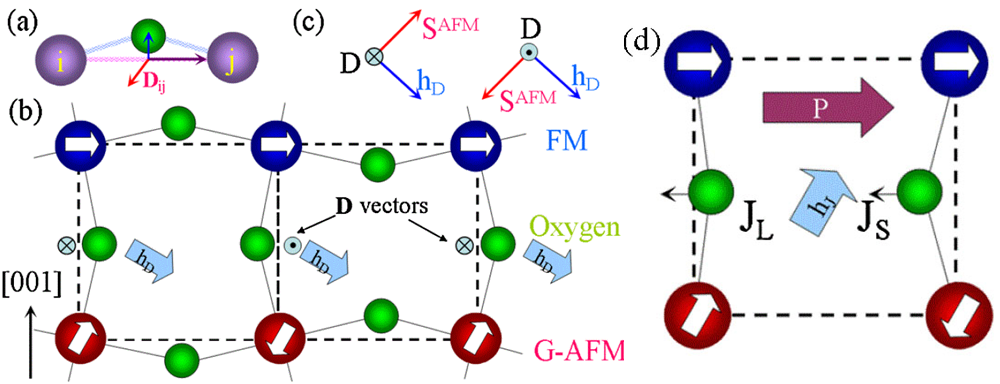}
\caption{(a) The relationship between the $M_i$-O-$M_j$ bond, oxygen displacement and $\vec{D}_{ij}$ vector. (b) Sketch of the interface between ferromagnetic and G-type antiferromagnetic perovskites, including the oxygen octahedral tilting. (c) The uniform $\vec{h}_{D}$ should be perpendicular to $\vec{S}^{\rm AFM}$ and $\vec{D}$. (d) Ferroelectric-polarization-driven asymmetric bond angles and modulated normal superexchange couplings at the interface. Reprinted figure with permission granted from S. Dong \textit{et al.}, \textit{Phys. Rev. Lett.} \textbf{103} (2009) 127201. Copyright \copyright (2009) by the American Physical Society, \href{http://dx.doi.org/10.1103/PhysRevLett.103.127201}{http://dx.doi.org/10.1103/PhysRevLett.103.127201}.} \label{f6}
\end{figure}

Several possible mechanisms have been proposed to understand the exchange bias effect. Extrinsic factors are often considered, such as interface roughness, spin canting near the interface, frozen interfacial and domain pinning.\cite{Malozemoff:Prb,Koon:Prl,kiwi:Apl,Schulthess:Prl,kiwi:Mmm} While the purely magnetic interactions framework stemming from traditional metallic magnetism appears incomplete to deal with the exchange bias effect unveiled in these magnetic oxides heterostructures, Dong \textit{et al.}\cite{Dong:Prl2} proposed two (related) mechanisms based on the Dzyaloshinskii-Moriya interaction and on the ferroelectric polarization, to understand the intrinsic exchange bias effect, in particular to understand the exchange bias effect in ferromagnetic/G-type antiferromagnetic oxides heterostructures. The latter mechanism is only active in those heterostructures that involve multiferroics.
The spin-spin interaction in perovskites can be described by a simplified Hamiltonian:
\begin{equation}
H = \sum_{<ij>}[J_{ij}\vec{S}_{i}\cdot \vec{S}_{j}+\vec{D}_{ij}\cdot (\vec{S}_{i}\times \vec{S}_{j})], \label{3}
\end{equation}
where $J_{ij}$ is the standard superexchange coupling between the nearest-neighbor spins, $\vec{D}_{ij}$ is the Dzyaloshinskii-Moriya interaction which arises from the spin-orbit coupling. In perovskites, $\vec{D}_{ij}$ is determined by the bending of the $M_i$-O-$M_j$ bond ($M$: metal cation) which is induced by the oxygen octahedral rotations and tilting. Moreover, the vector is perpendicular to the $M_i$-O-$M_j$ bond, as shown in Fig.~\ref{f6}(a). Since the tilting and rotations are collective, the nearest-neighbor oxygens in the same direction will move away from midpoint in opposite directions, namely, the nearest-neighbor displacements are staggered, as shown in Fig.~\ref{f6}(c). Thus, the $\vec{D}_{ij}$ vectors between nearest-neighbor bonds along the same direction are also staggered. It is noted that the spins in G-type antiferromagnetic S$^{\rm AFM}$ are also staggered. Combining these two staggered components $\vec{D}_{ij}$ and $\vec{S}^{\rm AFM}$ will give rise to a uniform Dzyaloshinskii-Moriya effect at the interface, as shown in Fig.~\ref{f6}(c)-(d), which can be described by an effective Hamiltonian:
\begin{equation}
H^{\rm interface}_{\rm DM} = \sum_{<ij>}\vec{D}_{ij}\cdot (\vec{S}_{i}^{\rm FM}\times \vec{S}_{j}^{\rm AFM})=-\vec{h}_{D}\cdot \sum_{i}\vec{S}_{i}^{\rm FM},
\label{4}
\end{equation}
where $\vec{h}_{D}=\vec{D}\times\vec{S}_{j}^{\rm AFM}$ can be regarded as the effective biased magnetic field which is fixed by the field-cooling process, then assumed to be frozen at low temperatures during the hysteresis loop measurement.

Furthermore, if one component has a ferroelectric polarization which induces a uniform displacement between cations and anions, the bond angles at the interface become no longer symmetric. Since the magnitude of normal superexchange coupling depends on the bond angle, the modulated bond angles induce staggered interfacial superexchange couplings, which are denoted as $J_{\rm L}$ and $J_{\rm S}$, as shown in Fig.~\ref{f6}(d). Once again, the staggered superexchange couplings at the interface will also induce a uniform bias field when in the presence of the G-type antiferromagnetic spin order, which can be described by:
\begin{equation}
H^{\rm interface}_{\rm DM} = \sum_{<ij>}\vec{J}_{ij}\cdot (\vec{S}_{i}^{\rm FM}\cdot \vec{S}_{j}^{\rm AFM})=-\vec{h}_{J}\cdot \sum_{i}\vec{S}_{i}^{\rm FM},
\label{5}
\end{equation}
where $\vec{h}_{J}=\frac{(J_{\rm S}-J_{\rm L})}{2}\vec{S}_{j}^{\rm AFM}$ comes from the modulation of superexchange $J$.

This model emphasizes the interactions between the lattice distortion and magnetism rather than rely on the existence of uncompensated antiferromagnetic moments anymore, which are often used in other models. Next, using the first principles theory, Dong \textit{et al.} chose the SrRuO$_3$/SrMnO$_3$ system to verify these two proposals for the exchange bias effect.\cite{Dong:Prb11.2}

\section{Ferroelectric field effect transistors \& tunnel junctions}

\subsection{Ferroelectric field effect transistors}
\begin{figure}
\centering
\includegraphics[width=\textwidth]{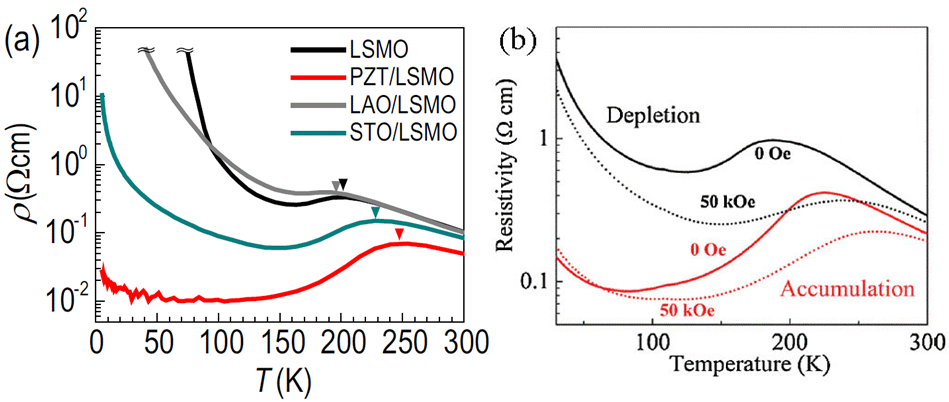}
\caption{(a) Transports properties of ultrathin La$_{0.8}$Sr$_{0.2}$MnO$_3$ films with different capping layers. The ferromagnetic Curie temperatures are denoted by triangles. Reprinted figure with permission granted from L. Jiang \textit{et al.}, \textit{Appl. Phys. Lett.} \textbf{101} (2012) 042902. Copyright \copyright (2012) by the American Institute of Physical, \href{http://dx.doi.org/10.1063/1.4738784}{http://dx.doi.org/10.1063/1.4738784}. (b) Resistivity as a function of temperature for the accumulation (red) and depletion (black) states at zero and at $50$ kOe magnetic field. Reprinted figure with permission granted from C. A. F. Vaz \textit{et al.}, \textit{Phys. Rev. Lett.} \textbf{104} (2010) 127202. Copyright \copyright (2010) by the American Physical Society, \href{http://dx.doi.org/10.1103/PhysRevLett.104.127202}{http://dx.doi.org/10.1103/PhysRevLett.104.127202}.} \label{f7}
\end{figure}

Besides affecting the magnetic properties, ferroelectric can also tune the transport properties. A ferroelectric field effect transistor, in analogy to the semiconductor counterpart, is basically composed of a controlling gate and a conductive channel. Here the gate is a ferroelectric film, while the channel is a metallic or semiconductor film. The conductivity of channel can be switched on and off by tuning the polarization of ferroelectric gate. As mentioned before, the magnetoelectric effect is prominent in the aforementioned La$_{0.8}$Sr$_{0.2}$MnO$_3$/PbZr$_{0.2}$Ti$_{0.8}$O$_3$ system, which originates from the charge-driven interfacial magnetic transition. Actually, there is also apparent variation of the in-plane conductivity when switching the polarization direction of PbZr$_{0.2}$Ti$_{0.8}$O$_3$.\cite{Vaz:Prl,Vaz:Apl,Jiang:Apl} The local carrier density at the interface is different for two polarization directions (accumulation state \textit{vs} depletion state), which drives a large change of the resistivity of La$_{0.8}$Sr$_{0.2}$MnO$_3$ film, as shown in Fig.~\ref{f7}(a). The bare La$_{0.8}$Sr$_{0.2}$MnO$_3$ film exhibits highly insulating behavior while the La$_{0.8}$Sr$_{0.2}$MnO$_3$/PbZr$_{0.2}$Ti$_{0.8}$O$_3$ heterostructure exhibits a metallic behavior over a wide range of temperature. The resistivity decreases by at least four orders of magnitude at $70$ K comparing with the bare La$_{0.8}$Sr$_{0.2}$MnO$_3$ film, and the ferromagnetic Curie temperature is about $250$ K. These prominent changes of transport properties is due to the ferroelectric layer since the system with non-ferroelectric capping layers (LaAlO$_3$ and SrTiO$_3$) show similar transport properties to the bare La$_{0.8}$Sr$_{0.2}$MnO$_3$ film. Fig.~\ref{f7}(b) shows the transport curves with/without external magnetic field for both the depletion and accumulation states of the La$_{0.8}$Sr$_{0.2}$MnO$_3$ when switching the PbZr$_{0.2}$Ti$_{0.8}$O$_3$ polarization. Significant changes of the resistivity value as well as the metal-insulator transition temperature are found.

\subsection{Ferroelectric tunnel junctions}
\begin{figure}
\centering
\includegraphics[width=\textwidth]{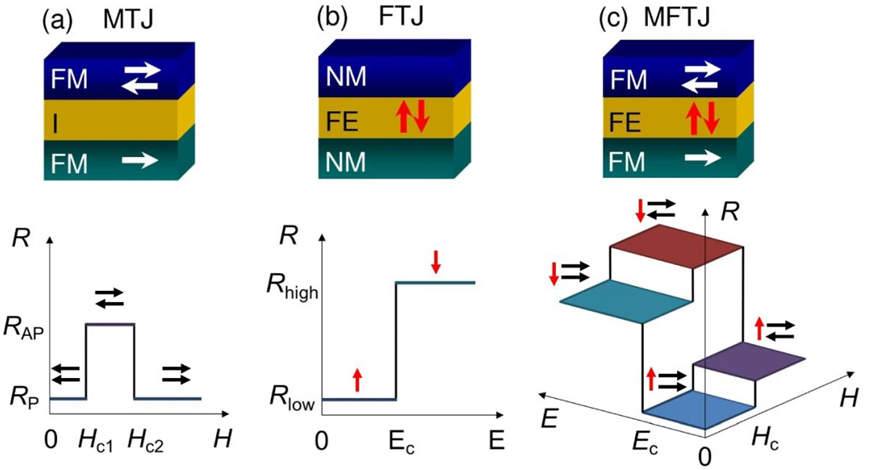}
\caption{Schematic view of tunnel junctions: (a) magnetic tunnel junction, (b) ferroelectric tunnel junction, (c) multiferroic tunnel junction. Bottom panels show the resistance response of these junctions to magnetic ($H$) and electric ($E$) fields. Different resistance states are demonstrated. Reprinted figure with permission granted from J. P. Velev \textit{et al.}, \textit{Phil. Trans. R. Soc. A} \textbf{369} (2011) 3069. Copyright \copyright (2011) by the Royal Society, \href{http://dx.doi.org/10.1098/rsta.2010.0344}{http://dx.doi.org/10.1098/rsta.2010.0344}.} \label{f8}
\end{figure}

Electron tunneling is a quantum-mechanical effect, where electrons can traverse a potential barrier that exceeds their kinetic energy. This phenomenon can be realized in a tunnel junction that consists of two metal electrodes separated by a nanometer-thick insulating barrier layer. Nowadays, significant interests in electron tunneling have been triggered by the applications of spintronics and numerous useful electronic devices. For example, in a magnetic tunnel junction consisting of two ferromagnetic metal layers separated by a thin insulating barrier, as shown in Fig.~\ref{f8}(a),\cite{Velev:Ptrsa} the tunnel current depends on the relative magnetization orientations of two ferromagnetic electrodes, which is known as tunneling magnetoresistance effect.\cite{julliere:Pla} The room-temperature large and reproducible tunneling magnetoresistance has been widely used in modern hard disks. In the following, we will introduce another two tunnel junctions, namely, ferroelectric tunnel junctions and multiferroic tunnel junctions, which have received enormous and continuous attentions recently.

Ferroelectric tunnel junctions take advantage of few-nanometer-thick ferroelectric as the tunnel barrier, as schematically depicted in Fig.~\ref{f8}(b). The basic idea of ferroelectric tunnel junctions was formulated in $1971$ by Esaki.\cite{Esaki:Itdb} In the last two decades, the progress of thin film technology made it realistic to fabricate ultrathin ferroelectric films while keeping their ferroelectricity,\cite{Bune:Nat,Tybell:Apl,Fong:Sci,Junquera:Nat,Tenne:Prl,Maksymovych:Prb,Tenne:Sci} which paved an exciting path to ferroelectric tunnel junctions.
\begin{figure}
\centering
\includegraphics[width=\textwidth]{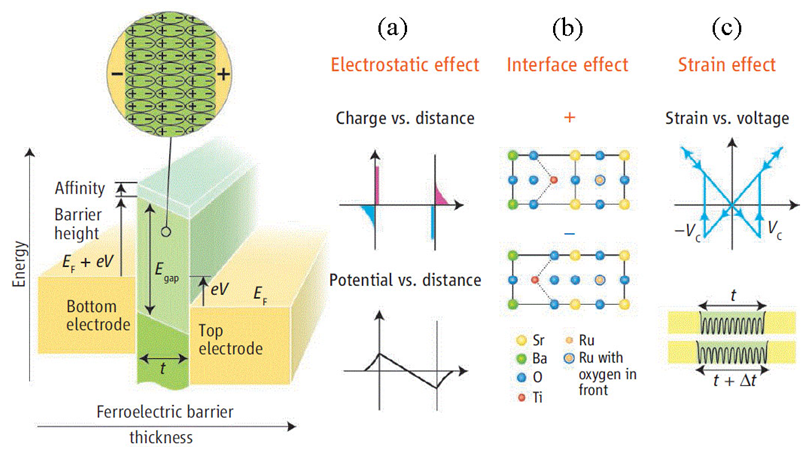}
\caption{Schematic diagram of a ferroelectric tunnel junction, which consists of two electrodes separated by a nanometer-thick ferroelectric barrier layer. $E_{\rm gap}$ is the energy gap. $E_{\rm F}$ is the Fermi energy, $V$ is the applied voltage, $V_{\rm c}$ is the coercive voltage, $t$ is the barrier thickness, and $\Delta_t$ is the thickness variation under an applied field. Mechanisms affecting tunneling in ferroelectric tunnel junctions: (a) the electrostatic potential across the junctions, (b) interface bonding and (c) strain effect associated with the piezoelectric response. Reprinted figure with permission granted from E. Y. Tsymbal \textit{et al.}, \textit{Science} \textbf{313} (2006) 181. Copyright \copyright (2006) by the American Association for the Advancement of Science, \href{http://dx.doi.org/10.1126/science.1126230}{http://dx.doi.org/10.1126/science.1126230}.}
\label{f9}
\end{figure}

In ferroelectric tunnel junctions, ferroelectric possess the spontaneous electric polarization that can be reversed by applying an external electric field, which may lead to a change in tunneling current, known as the tunneling electroresistance effect. As shown in Fig.~\ref{f9}, Tsymbal has summarized three critical elements that affect the interface transmission function and hence affect the tunneling electroresistance: (a) the electrostatic potential across the junctions, (b) interface bonding strength, and (c) strain associated with the piezoelectric response.\cite{Tsymbal:Sci} The electrostatic effect results from the incomplete screening of the polarization charges at the interface. This leads to an electrostatic potential that superimposes the contact potential in the tunnel junction, which can create an asymmetric potential profile with different electrodes. The interface effect comes from the ferroelectric displacements at the boundary between ferroelectric and electrodes.\cite{Fong:Prb} The atomic displacements will alter the orbital hybridizations at the interface as well as the transmission through it.\cite{Velev:Prl,Wortmann:Prb} The strain effect originates from the piezoelectric effect existed in all ferroelectrics. Distortions along the axis of junctions change the transport characteristics of the barrier such as the barrier thickness and hence affect the tunneling conductance.

\begin{figure}
\centering
\includegraphics[width=\textwidth]{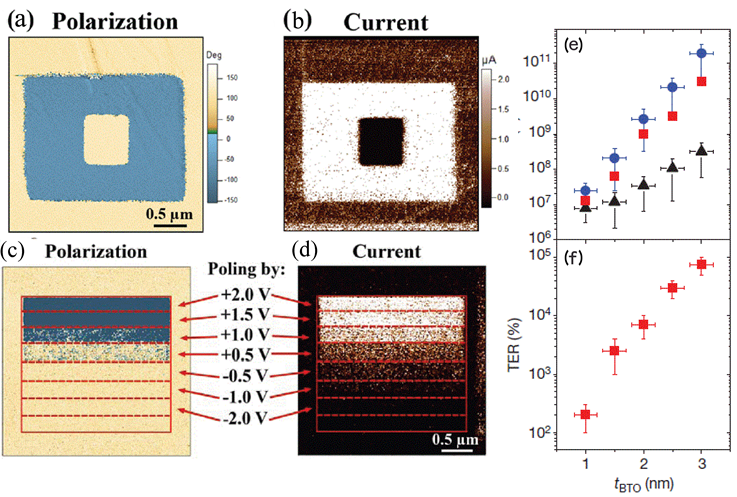}
\caption{(a-d) Tunneling electroresistance in BaTiO$_3$/SrRuO$_3$ ferroelectric tunnel junctions: (a) piezoresponse force microscopy image of a polarization pattern, (b) the corresponding tunneling current map measured by conducting atomic force microscopy. Spatially resolved correlation between (c) the onset of polarization reversal and (d) a change in electrical conductance. Blue: downward polarization; yellow: upward polarization. Bright: high current; dark: low current. Reprinted figure with permission granted from A. Gruverman \textit{et al.}, \textit{Nano Lett.} \textbf{9} (2009) 3539. Copyright \copyright (2009) by the American Chemistry Society, \href{http://dx.doi.org/10.1021/nl901754t}{http://dx.doi.org/10.1021/nl901754t}. (e-f) Observation of the giant tunneling electroresistance effect in ultrathin strained BaTiO$_3$ film. The relation between the thickness of BaTiO$_3$ and (e) resistance, (f) tunneling electroresistance effect in unpoled (red squares), and positively (black triangles) and negatively (blue circles) poled regions. Reprinted figure with permission granted from V. Garcia \textit{et al.}, \textit{Nature} \textbf{460} (2009) 81. Copyright \copyright (2009) by the Nature Publishing Group, \href{http://dx.doi.org/10.1038/nature08128}{http://dx.doi.org/10.1038/nature08128}.}
\label{f10}
\end{figure}

Experimentally, very large tunneling electroresistance effects were observed in BaTiO$_3$\cite{Gruverman:Nl,Garcia:Nat,Chanthbouala:Nn} and Pb$_{1-x}$Zr$_x$TiO$_3$\cite{Maksymovych:Sci,Pantel:Apl} ferroelectric thin films using scanning probe techniques. The comparison of Fig.~\ref{f10}(a)-(b) is a perfect correspondence of the spatial variation in tunnel currents to the polarization domains. It is clear seen that the area with downward polarization shows high conductivity while the area with upward polarization shows low conductivity. To further confirm the ferroelectric polarization induced resistive switching, the tunneling current was examined following the gradual change of polarization direction in the same area, as shown in Fig.~\ref{f10}(c)-(d). When the polarization direction undergoes reversal from the downward to upward, the tunnel current show a transition from high to low. Meanwhile, Garcia \textit{et al.}\cite{Garcia:Nat} revealed a giant electroresistance effect experimentally in BaTiO$_3$/La$_{0.67}$Sr$_{0.33}$MnO$_3$ ferroelectric tunnel junctions. They found that the resistance and the tunnel electroresistance effect ratio scales exponentially with the ferroelectric film thickness, reaching $\sim10000$\% and $\sim75000$\% at $2$ nm and $3$ nm, respectively, as shown in Fig.~\ref{f10}(e)-(f). These experimental findings unambiguously prove that the polarization can control the tunneling electroresistance effect, supporting earlier theoretical predictions.\cite{Zhuravlev:Prl,Kohlstedt:Prb}

\subsection{Multiferroic tunnel junctions}

When the magnetism of metallic electrodes is taken into account, the junctions can be considered as multiferroic tunnel devices, as shown in Fig.~\ref{f8}(c), which own the capability to control both the charge and spin tunneling via ferromagnetic and ferroelectric components in tunnel junctions. It is noted that resistance of such a multiferroic tunnel junction is significantly changed when the electric polarization of the barrier is reversed and/or when the magnetization of the electrodes is switched from parallel to antiparallel, rendering a four-state resistance device that the resistance can be controlled both by electric and magnetic fields.\cite{Zhuravlev:Apl}
\begin{figure}
\centering
\includegraphics[width=\textwidth]{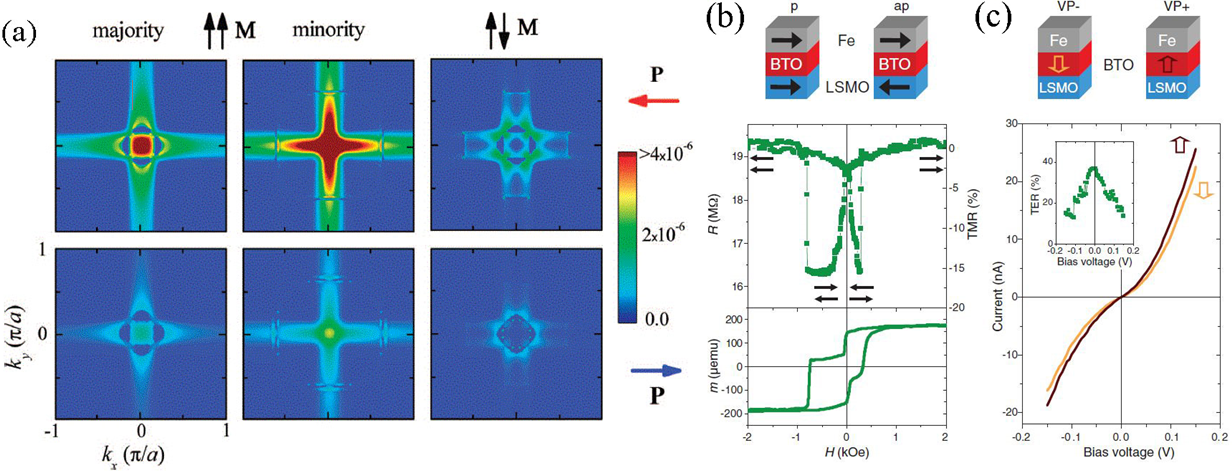}
\caption{(a) Transmission in the two-dimensional Brillouin zone of the SrRuO$_3$/BaTiO$_3$/SrRuO$_3$ multiferroic tunnel junctions. Calculated $k_{||}$-resolved transmission across the multiferroic tunnel junctions for different polarization $P$ of the barrier and magnetization $M$ of the electrodes. Reprinted figure with permission granted from J. P. Velev \textit{et al.}, \textit{Nano Lett.} \textbf{9} (2009) 427. Copyright \copyright (2009) by the American Chemistry Society, \href{http://dx.doi.org/10.1021/nl803318td}{http://dx.doi.org/10.1021/nl803318td}. (b-c) Magnetoresistive and electroresistive properties of Fe/BaTiO$_3$/La$_{0.67}$Sr$_{0.33}$MnO$_3$ multiferroic tunnel junctions. (b) Resistance (top) and magnetic moment (bottom) as a function of the magnetic field. (c) $I$-$V$ characteristics of the junction recorded at $4$ K after poling the ferroelectric BaTiO$_3$ barrier up and down. Reprinted figure with permission granted from V. Garcia \textit{et al.}, \textit{Science} \textbf{327} (2010) 1106. Copyright \copyright (2010) by the American Association for the Advancement of Science, \href{http://dx.doi.org/10.1126/science.1184028}{http://dx.doi.org/10.1126/science.1184028}.}   \label{f11}
\end{figure}

This four-state resistance was predicted in SrRuO$_3$/BaTiO$_3$/SrRuO$_3$ multiferroic tunnel junctions with asymmetric interfaces by using the first-principles theory.\cite{Velev:Nl} The tunneling magnetoresistance effect has the same origin as in ordinary magnetic tunnel junctions. When the two electrodes are in the parallel magnetic configuration, both spin channels (the majority-spin and minority-spin) contribute to the conductance, while in the antiparallel magnetic configuration, the conductance is strongly suppressed, as in the tunneling magnetoresistance devices. The tunneling electroresistance effect originates from the asymmetric interfaces. The asymmetric interfaces creates different polarization profiles when the ferroelectric polarization is switched. The transmission in the two-dimensional Brillouin zone of the SrRuO$_3$/BaTiO$_3$/SrRuO$_3$ multiferroic tunnel junction is shown in Fig.~\ref{f11}(a). The apparent differences in the transmission demonstrate the four resistance states can be controlled both by the electric and magnetic field.

Subsequently, the four-state resistance was realized experimentally.\cite{Garcia:Sci,Hambe:Afm,Yin:Fp,Pantel:Nm} Fig.~\ref{f11}(b)-(c) show the magnetoresistive and electroresistive properties of Fe/BaTiO$_3$/La$_{0.67}$Sr$_{0.33}$MnO$_3$ multiferroic tunnel junctions.\cite{Garcia:Sci} The tunneling magnetoresistance effect ($\sim-17$\%) was achieved by tuning the magnetization directions of Fe and La$_{\frac{2}{3}}$Sr$_{\frac{1}{3}}$MnO$_3$. By applying short voltage pulses of ¡À1 V, reversible changes ($\sim30$\%) of the tunneling resistance was observed, which linked to the variation of the barrier height. However, this multiferroic tunnel junctions can only work at low temperatures and the observed magnetoresistance is much smaller than the predicted value.\cite{Velev:Nl} Recently, Yin \textit{et al.}\cite{Yin:Fp,Yin:Jap} studied the La$_{0.7}$Ca$_{0.3}$MnO$_3$/(Ba, Sr)TiO$_3$ or La$_{0.7}$Ca$_{0.3}$MnO$_3$/(Ba, Sr)TiO$_3$ multiferroic tunnel junctions. Their multiferroic tunnel junctions also showed the four resistance states and an even large magnetoresistance effect (up to $\sim300$\%) was found. In addition, their multiferroic tunnel junctions may work at room temperature since (Ba, Sr)TiO$_3$ can sustain ferroelectricity up till room temperature.

By inserting a nanometer-thick La$_{0.5}$Ca$_{0.5}$MnO$_3$ interlayer into one interface of La$_{0.7}$Ca$_{0.3}$MnO$_3$/BaTiO$_3$ ferroelectric tunnel junctions, the tunneling electroresistance ratio was enhanced up to $\sim10000$\%.\cite{Yin:Nm} Here the La$_{0.5}$Ca$_{0.5}$MnO$_3$ acts as a magnetic supplementary of the ferroelectric barrier. In fact, Burton \textit{et al.} demonstrated that a magnetoelectric interaction between BaTiO$_3$ and La$_{1-x}$Sr$_x$MnO$_3$ magnetic electrode can generate a giant tunneling electroresistance effect due to the magnetic phase tuning of La$_{1-x}$Sr$_x$MnO$_3$.\cite{Burton:Prl} This idea have been realized experimentally recently in La$_{1-x}$Sr$_x$MnO$_3$/PZT heterostructures.\cite{Jiang:Nl} The ferroelectric-induced phase modulation at the heterointerface ultimately results in an enhanced electroresistance effect.

Another type of multiferroic tunnel junctions employs a single phase multiferroic as the tunnel barrier. Gajek \textit{et al.} first demonstrated a La$_{\frac{2}{3}}$Sr$_{\frac{1}{3}}$MnO$_3$/La$_{0.1}$Bi$_{0.9}$MnO$_3$/Au multiferroic tunnel junctions in which the tunneling resistance was controlled by both electric and magnetic fields.\cite{Gajek:Nm} In this multiferroic tunnel junctions, the tunneling magnetoresistance effect arises from the spin filtering effect and the tunneling electroresistance effect is related to the barrier potential change when its polarization is reversed. The main problem of this type multiferroic tunnel junctions is that the single phase multiferroic materials with stable ferromagnetic and ferroelectric orders are rare, since most multiferroic materials show antiferromagnetism. Besides, most of single phase multiferroic materials show low magnetic ordering temperatures. Despite low temperatures and low magnetoresistance in this type of multiferroic tunnel junctions, the discovery of four logical states is indeed an important progress to design and realize the applications of multiferroic tunnel junctions in the future.

\section{Summary and Perspective}

In this review, we have briefly introduced some recent theoretical and experimental progresses on controlling the magnetism and transport in oxide heterostructures via ferroelectric polarization. We mainly focused on the interfacial magnetoelectric effects and magneto- and electro-resistance effects. Regarding the interfacial magnetoelectric effects, there are several ways to tuning the interfacial magnetism, including the magnitude of magnetization, easy axis, magnetic phases, as well as exchange bias. For the transport effects, there are giant modulations of resistance in ferroelectric or multiferroic junctions associated with polarization switching.

The magnetoelectric effects in oxide heterostructures are very attractive due to both the fundamental scientific interests and promising technological potential. These effects provide an alternative route to explore the fascinating physics of correlated electronic materials and open the door to access faster and energy-efficient quantum electronic devices. In the current stage, there remain many scientific and technical issues to be solved, this fast developing field is full of challenges and opportunities.

\section*{Acknowledgments}
Work was supported by the Natural Science Foundation of China (grant Nos. 51322206 and 11274060) and the 973 Projects of China (grant No. 2011CB922101)

\bibliographystyle{ws-mplb}
\bibliography{ref}
\end{document}